# Rashba-Dresselhaus spin-orbit coupling and polarization-coupled luminescence in an organic single crystal microcavity

Reo Ohkura[1,†], Takaya Inukai[1,†], Shun Takahashi[1], Hitoshi Mizuno[2], Masaaki Nakayama[3], Yohei Yamamoto[4], and Kenichi Yamashita[1*]

[1]*Faculty of Electrical Engineering and Electronics, Kyoto Institute of Technology, Matsugasaki, Sakyo-ku, Kyoto 606-8585, Japan*

[2]*Graduate School of Materials Science, Nara Institute of Science and Technology, 8916-5 Takayama, Ikoma, Nara 630-0192, Japan*

[3]*Department of Applied Physics and Electronics, Graduate School of Engineering, Osaka City University[‡], Sugimoto, Sumiyoshi-ku, Osaka 558-8585, Japan*

[4]*Department of Materials Science and Tsukuba Research Center for Energy Materials Science (TREMS), Institute of Pure and Applied Sciences, University of Tsukuba, 1-1-1 Tennodai, Tsukuba, Ibaraki 305-8573, Japan*

[†]These authors equally contributed.

[‡]Merged with Osaka Prefecture University to form Osaka Metropolitan University.

*e-mail: yamasita@kit.ac.jp

Spin-orbit coupling (SOC) of light plays a fundamental photophysics that is important for various fields such as materials science, optics, and quantum technology, contributing to the elucidation of new physical phenomena and the development of innovative applications. In this study, we investigate the impact of SOC in a microcavity system using the highly oriented molecular crystal. The unique molecular alignment of our crystal creates substantial optical anisotropy, enabling the observation of significant SOC effects within a microcavity form. Through angle-resolved photoluminescence measurements and theoretical calculations, the presence of Rashba-Dresselhaus (RD) SOC in the lower branch of polariton modes is revealed. We have observed for the first time polarization-coupled emission from polariton modes due to the RD-SOC effect in a microcavity with a medium having both strong light-matter coupling and strong optical anisotropy. Theoretical investigations further elucidate the intricate interplay between the RD-SOC effect and anisotropic light-matter coupling, leading to the emergence of both circularly and diagonally polarized mode splittings. This study not only advances our understanding of optical SOC in microcavities but also highlights the potential of highly oriented molecular crystals in manipulating SOC effects without external electric or magnetic fields. These findings offer greatly promising platforms for developing topological photonics and quantum technologies.

## Introduction

Spin-orbit coupling (SOC) plays a crucial role in determining the electronic properties and material characteristics[1,2]. For example, significant SOCs based on Rashba and Dresselhaus (RD) effects are induced in solid-state systems with broken inversion symmetry[3–5]. These effects are involved in realizations of new physical phenomena and devices such as spintronics and topological insulators[6]. On the other hand, SOC in the optical regime refers to interaction between spin and orbital angular momentums of light in media with significant refractive index distribution or optical anisotropy[7–9]. The optical SOCs cause propagations of different optical spin states (circular polarization states) to bend in opposite directions perpendicular to the propagation direction[10]. This effect is known as the optical spin Hall effect and has attracted attention in the fields of topological photonics and spintronics.

It is known that optical microcavity systems also undergo SOC effects caused by the transverse electric (TE) – transverse magnetic (TM) splitting[10,11] and/or the optical anisotropy of microcavity material[12–14]. These types of SOCs cause effective magnetic fields for the optical spin states and are represented by artificial gauge fields[15,16]. When the inversion symmetry is broken, the microcavity system can be described by RD-type Hamiltonian, leading to the splitting of an optical mode in the in-plane wavevector direction. The split optical modes are spin polarized and show circularly polarized luminescence (CPL) with opposite helicities[17,18].

When microcavity system is in a coupling regime in terms of light-matter interaction, the RD-type SOC (RD-SOC) acts on the polaritonic modes instead of the photonic cavity modes. In a recent study, room-temperature (RT) Bose-Einstein condensation of the spin-polarized polariton modes has been demonstrated[17,18], which is promising as a fundamental photophysics for various quantum applications. Additionally, as the polariton state is hybrid states of light and matter, the fundamental quantum properties of the spin-polarized polariton states are attracted also for manipulating spin momentum transfer between the photonic and electronic states[19]. However, active media possessing both strong light-matter coupling and strong optical anisotropy are limited. While liquid crystal materials exhibit a significant anisotropy inducing inversion symmetry breaking, it is required that organic dyes or lead halide perovskites is introduced as active media for achieving strong coupling[17–19]. External voltage application is also needed, which may reduce oscillator strength of exciton and weaken the coupling strength with cavity mode. The RD-SOC in a microcavity using an organic crystal having an anisotropic dielectric tensor has been demonstrated in a recent study through transmission measurements[16,20], but that study has not demonstrated circularly polarized light emission.

In this study, we investigate the optical SOC in a microcavity utilizing a highly oriented molecular crystal, BP1T-CN, which has been demonstrated to undergo light-matter strong coupling and polariton condensation at RT in a microcavity form[21–23]. Angle-resolved photoluminescence (ARPL) measurements utilizing a Fourier imaging spectroscopy setup clearly reveal RD-SOC in the lower branch of polariton modes. The Rashba constant is determined to be 1.62 nm·eV that is notably larger than that observed in previous studies[16]. Furthermore, for the first time, the circularly polarized component of the emission signal was clearly observed in a microcavity with a medium having both strong light-matter coupling and strong optical anisotropy. We also perform a theoretical calculation for energy dispersion and polarization states of the split polariton modes, revealing intricate behaviors arising from the interplay between the RD-SOC effect and the anisotropic light-matter coupling. These results provide valuable insights not only for studying topological photonics but also for constructing quantum technologies involving spin interactions between photonic and electronic systems.

## Results and discussions

The focus of this study about the microcavity material is on 2,5-bis(4'-cyanobipheny-4-yl) thiophene, also known as BP1T-CN [see Fig. 1(a)][24]. The BP1T-CN molecule exhibits a substantial transition dipole moment along its long molecular axis. In a form of platelet single crystal, all molecules align unidirectionally in a triclinic crystal system [see Fig. 1(b)]. This pronounced molecular orientation results in significantly large optical anisotropy, as depicted in Fig. S1 of the Supplementary Information. The anisotropic refractive index spectra of bare BP1T-CN crystal are shown in Fig. S2 of the Supplementary Information, which are determined by projecting the index ellipsoid of crystal onto the *xy*-plane. Importantly, as shown in Fig. S3 in the Supplementary information, the principal optical axis of refractive index ellipsoid is at an angle of about 22 ° with the surface plane of crystal (*xy*-plane)[24], which is a key factor to break the inversion symmetry of polaritonic modes within the cavity. We have previously confirmed that microcavities utilizing BP1T-CN crystals [see Fig. 1(c)] demonstrate strong light-matter coupling at RT[22,25], leading to the formation of upper-polariton (UP) and lower-polariton (LP) modes [see Fig. 1(d)]. The highly anisotropic optical property of the BP1T-CN crystal induces a separation between horizontally and vertically polarized LP modes at a wavenumber vector of $\boldsymbol{k} = 0$ [i.e. X-Y splitting, see Fig. 1(d)][12]. Additionally, at high $|\boldsymbol{k}|$ regions, the LP mode exhibits TE-TM splitting, arising from the difference in effective refractive indices of the TE and TM modes, which occurs even in an isotropic microcavity.

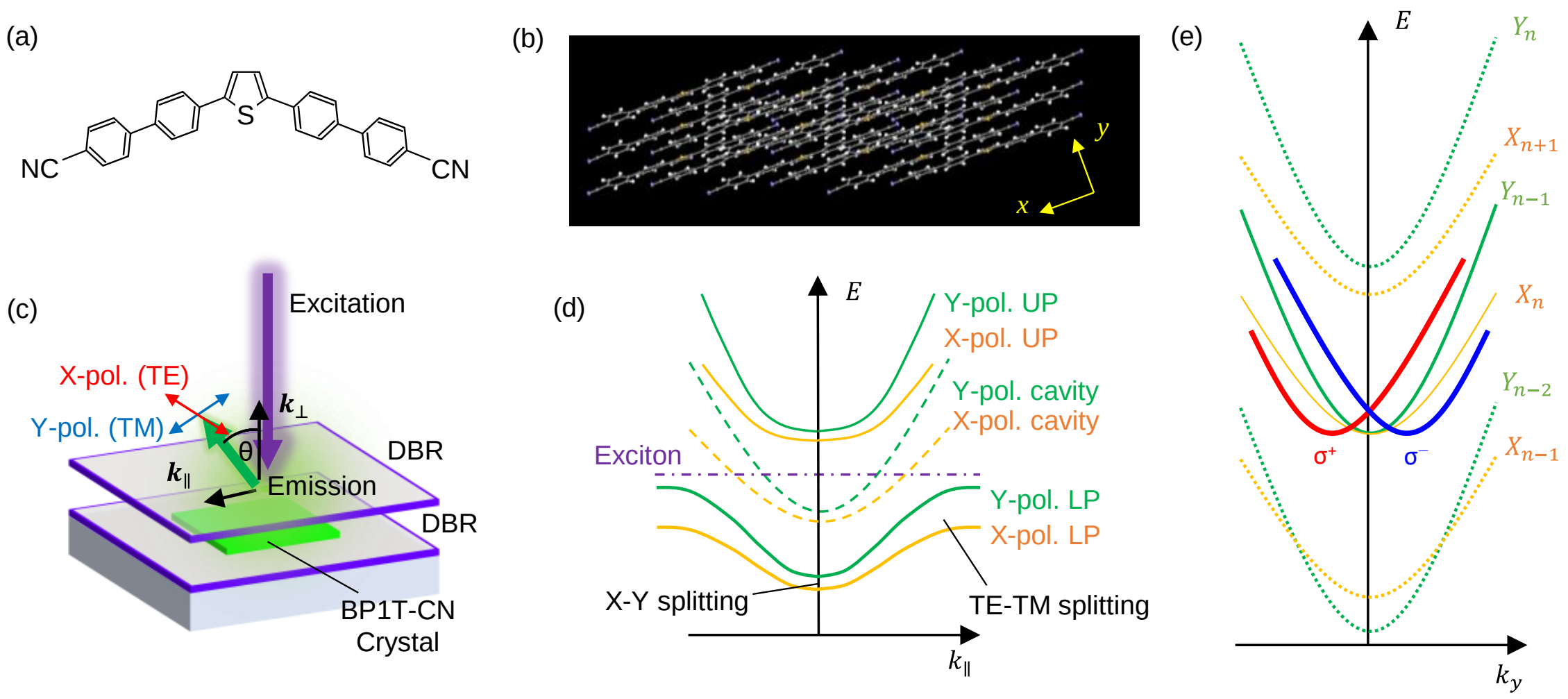

**Figure 1.** BP1T-CN microcavity. (a) Molecular formula of BP1T-CN. (b) A schematic of Molecular alignment in BP1T-CN crystal. (c) A schematic depiction for BP1T-CN microcavity and a configuration of ARPL. (d) An overview of dispersion curves of polariton modes with X-Y and TE-TM splittings. (e) Dispersion curves undergoing RD-SOC effect. When the when the two opposite polarized modes of opposite parity are tuned near resonance, the LP mode is split along $k_y$ direction with circular polarizations.

By leveraging the substantial X-Y splitting in a microcavity with BP1T-CN crystal, it becomes possible to achieve energy resonance at $\boldsymbol{k} = 0$ between two orthogonally polarized polariton modes at certain cavity length. As shown in Fig. 1(e), previous studies have demonstrated the emergence of RD-SOC when the two polarized modes of opposite parity ($X_n$ and $Y_{n-1}$) are brought into proximity to resonance[15]. The effective Hamiltonian, $H(\boldsymbol{k})$, governing this SOC effect, can be derived from Maxwell's equations in the microcavity structure. In this framework, the X-Y splitting and TE-TM splitting are considered as effective magnetic fields acting on the optical spinor within the in-plane momentum ($k_x k_y$) space.

$$H(\boldsymbol{k}) = \begin{bmatrix} \frac{\hbar^2 k_x^2}{2m_x} + \frac{\hbar^2 k_y^2}{2m_y} - 2\alpha k_y & \beta_0 + \beta_1 \boldsymbol{k}^2 e^{2i\varphi} \\ \beta_0 + \beta_1 \boldsymbol{k}^2 e^{-2i\varphi} & \frac{\hbar^2 k_x^2}{2m_x} + \frac{\hbar^2 k_y^2}{2m_y} + 2\alpha k_y \end{bmatrix} \quad (1)$$

Here, $m_x$ ($m_y$) is the effective mass of lower polariton mode along $x$ ($y$) direction. $\beta_0$ and $\beta_1$ denote the magnitudes of X-Y splitting and TE-TM splitting, respectively. $\alpha$ represents the RD splitting along $k_y$ direction. Note here that the contributions of Rashba and Dresselhaus couplings are equal in this case[15,16].

The RD-SOC effect in BP1T-CN single crystal microcavity can be evaluated through ARPL measurements. Transmission spectra of high-reflectivity distributed-Bragg reflectors (DBRs) fabricated by rf-magnetron sputtering method are shown in Fig. S4 in the Supplementary information. We have used a Fourier-imaging setup for the ARPL measurements (see Fig. S5 in the Supplementary information). Figures 2(a) and 2(b) display unpolarized ARPL results obtained from a microcavity sample with a platelet BP1T-CN single crystal. Figure S6 in the Supplementary information exhibits an optical microscopic image of the crystal having ~40 × 25 μm$^2$ in a dimension and ~365 nm in thickness. The transition dipole moment within the crystal projected onto the *xy* plane lies along the *x* direction, as illustrated in Fig. 1(b). In the colormap showing the photoluminescence (PL) dispersion at the $k_x$ = 0 plane [see Fig. 2(a)], distinct splitting behaviors of the LP modes along the $k_y$ direction are observed in an energy range of ~2.445 eV. This observation can be attributed to the RD splitting resulting from the coupling of $X_4$ and $Y_3$ modes. Another splitting mode appearing on the low energy side (~2.415 eV) is a ghost signal from a site with slightly different crystal thicknesses and exhibits the same behavior as the main signal. Conversely, the PL dispersion at the $k_y$ = 0 plane [see Fig. 2(b)] does not exhibit any splitting at $k_x$ = 0, neither in the wavenumber nor energy directions, whereas the TE-TM splitting is observed at high $k_x$ values. It should be noted that the weakened ARPL signal in the negative $k_x$ region arises from the tilt in the molecular orientation of BP1T-CN (i.e., direction of transition dipole moment) along the *z* direction.

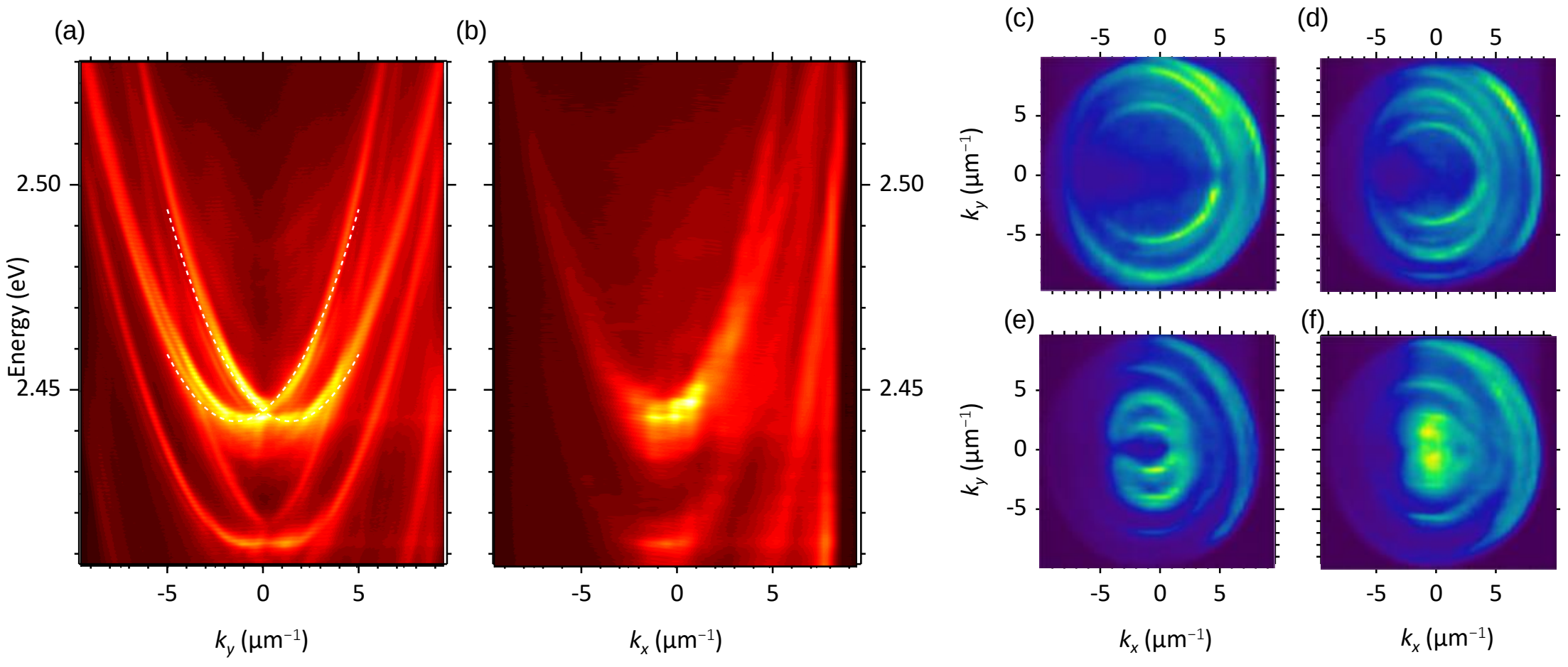

**Figure 2.** ARPL characterization of dispersion surface using Fourier imaging method. (a, b) Dispersion curves on the energy - $k_y$ (a) and energy - $k_x$ (b) planes. (c – f) Dispersion ellipsoids on isoenergetic planes at 2.512 (c), 2.483 (d), 2.455 (e), and 2.439 eV (f).

To gain a more understanding of the three-dimensional structure of this dispersive surface, we have included two Supplementary videos, S1 and S2, which show dispersions as the in-plane azimuthal angle continuously changes and its isoenergetic cross-sections in the $k_x k_y$ plane.

Snapshots of the isoenergetic cross-sections at four different energies are presented in Figs. 2(c) to 2(f). At higher energy [see Fig. 2(c)], two elliptical modes corresponding to the TE and TM modes can be observed within the range determined by the numerical aperture (NA) of objective used (~0.75). The ellipticity arises from the anisotropy in the light-matter coupling strength between the two orthogonally polarized modes; the TE mode (outer one) exhibits a stronger coupling compared to the TM mode[12], as depicted in Fig. 2(d). As the energy decreases, the mode profile undergoes topological phase changes; (i) the TE and TM modes intersect and give rise to two intersecting circular modes [see Fig. 2(e)] and (ii) these two modes split along the $k_y$ direction, demonstrating the RD effect [see Fig. 2(f)]. The nearly degenerate nature of the two modes at $\boldsymbol{k} = 0$ indicates that the original $X_4$ and $Y_3$ LP modes are almost fully resonant ($\beta_0 \sim 0$). The RD parameter has been evaluated as $\alpha \sim 1.77 – 1.88$ nm·eV (see Note S1 and Fig. S7 in the Supplementary information), which is significantly larger than the recent finding for $\beta$-phase single-crystalline perylene (< 1 nm·eV)[16].

Figure 3 shows the dispersion characteristics for Stokes parameters, $S_1$, $S_2$, and $S_3$. In the experiments, we clearly see distinct circularly polarized component in the luminescence signals from the LP modes split by RD-SOC effect, as shown in Fig. 3(c). Specifically, the mode split in the negative (positive) $k_y$ direction results in right (left) hand circularly polarized emission, $\sigma_+$ ($\sigma_-$). This result represents the first successful demonstration of polaritonic CPL from a microcavity utilizing a single active medium with the RD-SOC effect. While the bare BP1T-CN crystal originally emits linearly X-polarized light[12,22], the RD-SOC effect alters the vacuum field within the cavity, leading to the splitting of light into $\sigma_+$ and $\sigma_-$ polarized components.

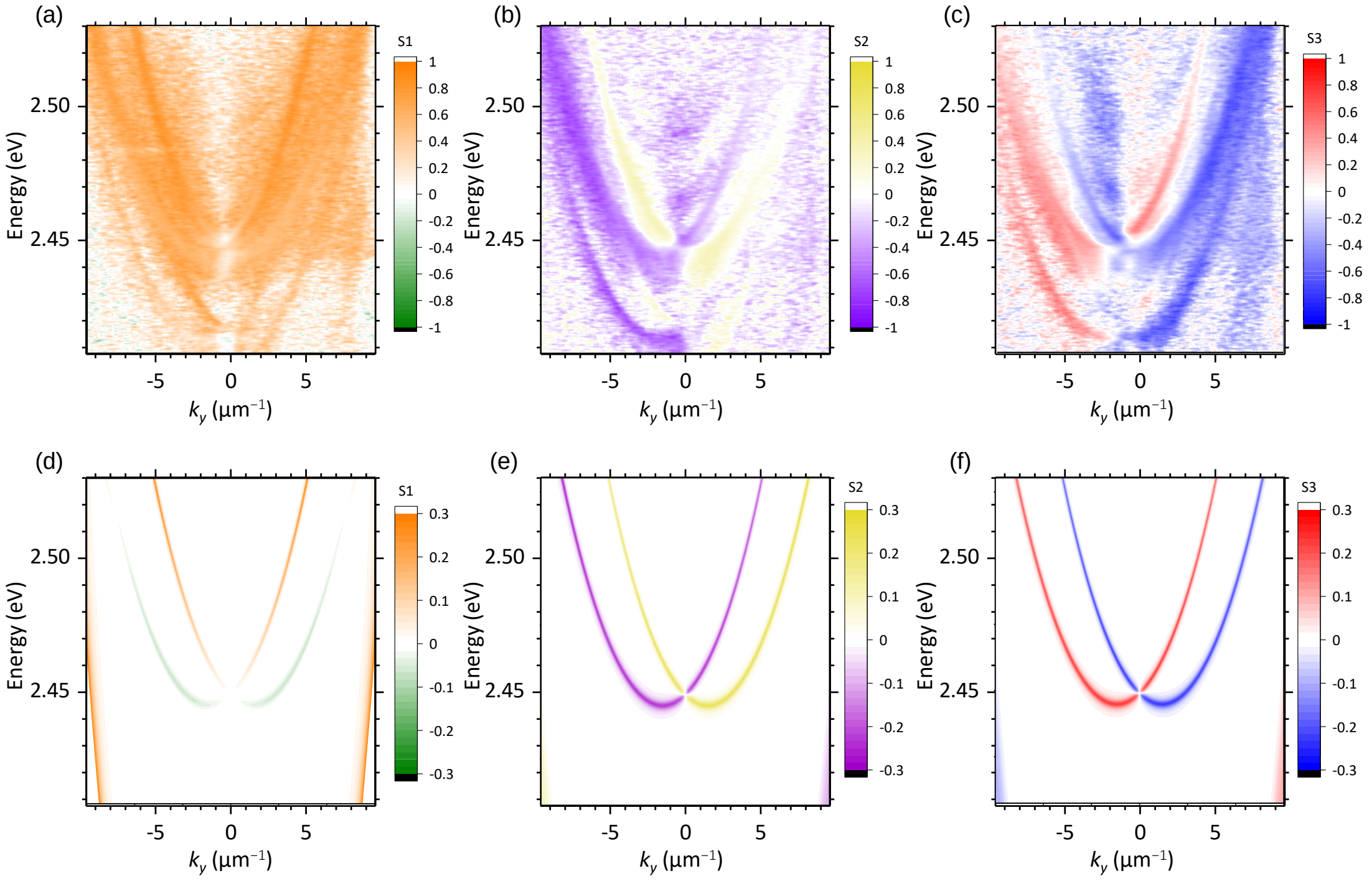

**Figure 3.** Colormaps for Stokes parameters of LP modes in BP1T-CN microcavity. (a – c) $S_1$ (a), $S_2$ (b), and $S_3$ (c) obtained from ARPL measurements. (d – f) $S_1$ (d), $S_2$ (e), and $S_3$ (f) calculated using Berreman 4 × 4 matrix method.

We investigate the detailed polarization state of LP mode emission associated with the RD-SOC effect utilizing a theoretical calculation. Figures 3(d) – 3(f) exhibit Stokes parameters for reflectivity spectrum calculated with Berreman 4 × 4 matrix method. Note that the model of microcavity used for this calculation is not perfectly the same with the experimental device (see Materials and method); in particular, the imaginary part of refractive index has been assumed to be zero and the index dispersion of BP1T-CN crystal has not been included in the calculations. However, as we focus only on a limited energy range of ~2.4 – 2.5 eV, this calculation is sufficient to understand the polarization characteristics attributed to the RD-SOC effect. In Fig. 3(d), we find that the calculated $S_1$ of two LP modes split by the RD-SOC effect exhibit sign reversal between the positive and negative wavenumber regions. In the experimental results obtained from the PL measurements [see Fig. 3(a)], on the other hand, $S_1$ for the LP modes remains in positive values in all observed wavenumber regions. This is because the original emission from BP1T-CN molecule is almost polarized to the X direction in this measurement configuration and thus there is no Y-polarized component in the emission signal. $S_2$, which represents a linear polarization degree between the diagonal (45 º) and anti-diagonal (135 º) polarizations, exhibits opposite signs to each other between the two LP modes [see Fig. 3(e)]. This feature can be observed in the experimental results shown in Fig. 3(b), as similarly to a recent result of twisted nematic liquid crystal microcavity[19]. The good agreement between the experiment and calculation results is found also for $S_3$ [see Fig. 3(f)], revealing that the BP1T-CN microcavity is a good platform as a RD-SOC polariton state.

The polarization state reconstructed through the RD-SOC effect in the BP1T-CN microcavity can be characterized as an almost ideal pure state in the quantum theory perspective. Figure 4(a) shows a contour plot of $S_0^2$ $(= S_1^2 + S_2^2 + S_3^2)$ in the vicinity of the energy minimum of the LP modes. The outer LP mode exhibits $S_0^2$ values approaching unity, revealing that the polarization state lies on the surface of Poincaré sphere. On the other hand, $S_0^2$ values for the inner LP mode are marginally degraded (~0.8), probably due to the imperfection of microcavity. Notably, at $k_y$ = 0, $S_0^2$ value diminishes nearly to zero, revealing a perfectly unpolarized state (mixed state). To illustrate the specialty of this finding, we present a series of computational outcomes for $S_0^2$ at various detuning of the original $X_4$ and $Y_3$ LP modes [see Figs. 4(b) – 4(d)]. By changing $\theta$ (the angle between the optical principal axis and the *xy*-plane) from 22.5 ° by approximately ±0.5 ° and slightly modulating the detuning in both the positive and negative directions, we observed that the $S_0^2$ value no longer remains zero at $k_y$ = 0. The clear emergence of a polarization singularity in our experimental result clearly indicates that the detuning of the original $X_4$ and $Y_3$ LP modes is effectively negligible. This surprising result is caused by a peculiar polarization characteristic of the BP1T-CN microcavity.

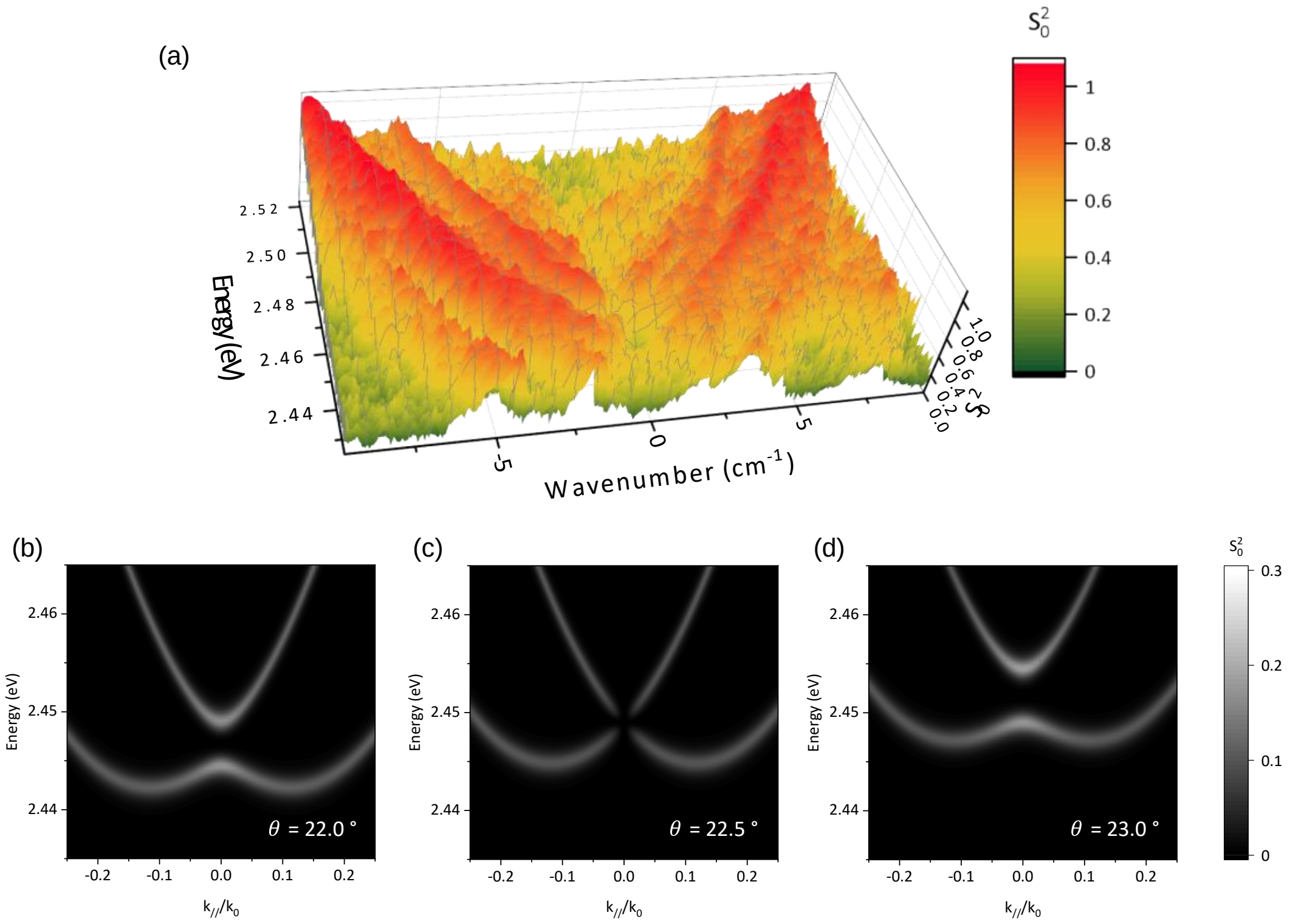

**Figure 4.** Degree of polarization of LP modes split by RD-SOC. (a) A contour plot of experimentally obtained $S_0^2$ in the vicinity of the energy minimum of the LP mode. (b – d) $S_0^2$ at $\theta$ = 22.0 ° (b), 22.5 ° (c), and 23.0 ° (d) calculated using Berreman 4 × 4 matrix method.

Finally, we present a comprehensive elucidation of the polarization states within the BP1T-CN microcavity, offering valuable insights into the topological characteristics of polarization state [19,20,26]. As depicted in Fig. 5, isoenergetic cross-sections of dispersion surfaces for $S_3$ are calculated at three distinct crystal thicknesses. In this calculation, a fixed coordinate system is employed to describe the experimental configuration, whereas the sample undergoes a complete 360° rotation along the z-axis. Complete datasets encompassing $S_1$ and $S_2$ are shown in Figs. S8, S9, and S10 in the Supplementary information.

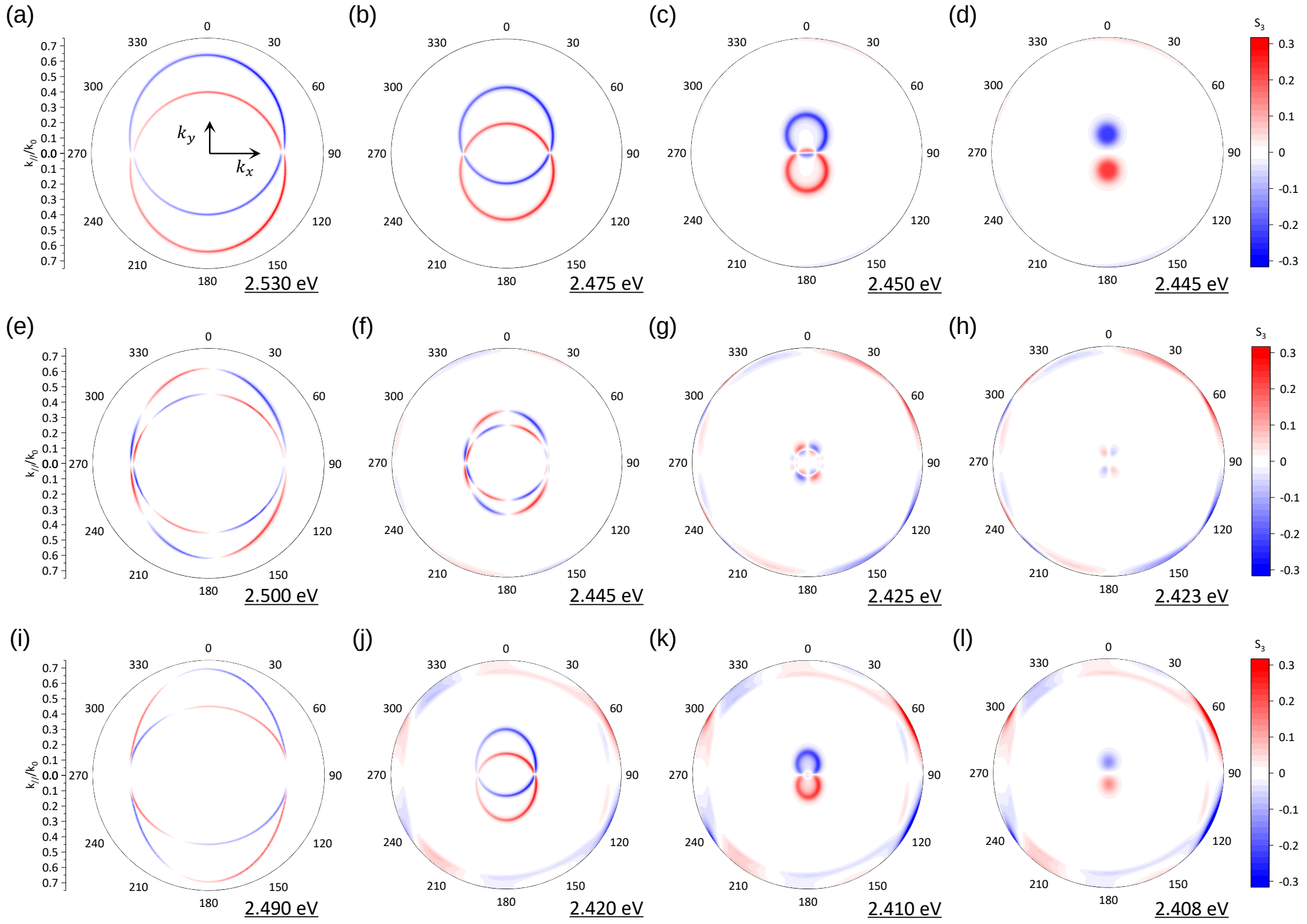

**Figure 5.** Colormaps of $S_3$ in $k_xk_y$-plane calculated using Berreman 4 × 4 matrix method. Energy cross-sections calculated at BP1T-CN crystal thicknesses of 365 nm (a – d), 730 nm (e – h), and 1,095 nm (i – l), where $X_4 - Y_3$, $X_8 - Y_6$, and $X_{12} - Y_9$ resonances emerge, respectively. Energy for which calculation has been performed is shown in each figure.

At a crystal thickness of 365 nm, mirroring the experimental conditions, the dispersion profile of $S_1$ exhibits a distinct double elliptical structure at higher energy [~2.530 eV, see Fig. S8(a) in the Supplementary information]. Notably, the outer mode reveals a larger elliptical geometry in comparison to the inner mode. This observation stems from the anisotropic coupling between light and matter, influencing the formation of polariton modes, as shown in Figs. 2(c). Due to the RD-SOC effect, both $S_2$ and $S_3$ also manifest non-zero components, as exhibited in Figs. S8(b) and S8(c), respectively. Importantly, we have verified that these non-zero polarizations of $S_2$ and $S_3$ arise exclusively when the optical principal axis is tilted relative to the $z$ axis (i.e. when

$\theta \neq 0°$).

As shown in Figs. 5(a) and 5(b), the dispersion profile of $S_3$ undergoes a notable transformation: a transition from two concentric ellipsoids with singular winding in polarization to two intersecting ellipsoids with single polarities ($\sigma_+$ and $\sigma_-$). Upon further reduction in energy, the two ellipsoids distinctly diverge along the $k_y$ axis and converge towards local minima with the opposite polarities [see Figs. 5(c) and 5(d)]. These topological shifts in polarization features are typical phenomena associating with the RD-SOC effect with the $X_n$ and $Y_{n-1}$ resonance[15,16].

Nevertheless, the inherent topological properties of polarization undergo a pronounce change upon the manipulation of crystal thickness. When the thickness increases to 1,095 nm where the $X_{12}$ and $Y_9$ resonance should emerge, the behavior of two ellipsoids undergoes a dual winding in polarization [see Fig. 5(i)]. This result diverges from the scenario observed with the $X_4$ and $Y_3$ resonance, which exhibits a singular winding in polarization [see Fig. 5(a)]. It is noteworthy that both of these two cases satisfy the requisite conditions of resonance between the modes of differing parity, enabling the splitting into two modes with opposite polarities [see Figs. 5(d) and 5(l)]. However, the difference in the mode number difference (one or three) results in a variation in the topological charge of polarization [see Figs. 5(a) and 5(i)]. As another intriguing thing, the profiles of $S_2$ and $S_3$ display symmetrical characteristics relative to the $k_y$ axis [see Figs. S10(e) – S10(l) in the Supplementary information]. This finding indicates that the circular and the diagonal polarization mode splittings arise simultaneously with a complementary nature.

At the crystal thickness of 730 nm, a distinct deviation arises in the context of RD-SOC due to the establishment of resonance between modes of identical parity ($X_8$ and $Y_6$). We found in this case that more intriguing topological properties appear, as shown in Figs. 5(e) – 5(h) and S9 in the Supplementary information. Figure 5(e) exhibits that, within a high-energy range, the two ellipsoids corresponding to $S_3$ shows a triple winding in polarization. Interestingly, as the energy decreases, the number of polarization windings increases [see Fig. 5(f)], which is opposite trend to the case of 1,095 nm. Also we observe no transition from two concentric ellipsoids to two intersecting ellipsoids [see Fig. 5(g)]. Finally, a quadrupole-like distribution of polarity appears for both $S_2$ and $S_3$ [see Fig. 5(h)], implying that it is interesting to explore the requisite for RD-SOC from the aspect of topological property on polarization. These findings hold significant promise for the exploration of novel topological polariton physics.

## Conclusions

This study provided a comprehensive investigation into the optical SOC effects in a microcavity system utilizing the highly oriented molecular crystal, BP1T-CN. The unique molecular alignment of BP1T-CN, which induces substantial optical anisotropy, enabled the realization of significant SOC effects in a microcavity form. Through ARPL measurements and theoretical calculations, we revealed the presence of RD-SOC in the lower branch of polariton modes. The Rashba constant was determined to be significantly larger than those observed in previous studies, indicating the effectiveness of BP1T-CN for inducing strong SOC. Importantly, we observed for the first time polarization-coupled emission from polariton modes due to the RD-SOC effect in a microcavity with a medium having both strong light-matter coupling and strong optical anisotropy, providing valuable insights for the manipulation of polariton states for potential quantum technology applications. Furthermore, theoretical investigations for the polarization states revealed intricate behaviors arising from the interplay between the RD-SOC effect and the anisotropic light-matter coupling. The emergence of both circularly-polarized and diagonally-polarized mode splittings demonstrates the richness of the optical SOC phenomena in this system. The obtained results contribute not only to the understanding of the fundamental physics of optical SOC in microcavities but also open up possibilities for the exploration of novel topological polariton physics.

Most importantly, this study sheds light on the potential of utilizing highly oriented molecular crystals like BP1T-CN to engineer and manipulate SOC effects in microcavity systems without any external electric and magnetic fields. As a future work, the observation of RT polariton lasing with circular polarization feature is attractive, although currently we have not been able to lower the condensation threshold below the damage threshold to the crystals. The strong light-matter coupling, along with the pronounced SOC effects, makes organic microcavities promising platforms for advancing the fields of topological photonics and quantum technologies involving spin interactions between photonic and electronic systems.

## Materials and methods

### *Molecular crystal growth*

Powdery BP1T-CNwas purchased from Sumitomo Seika Chemicals and used as received. The BP1T-CN crystal was grown by the sublimation-and-recrystallization method in a glass cylinder filled and purged with the dried nitrogen gas. The temperature inside the cylinder was spatially graded using a couple of band heaters; the temperatures for sublimation and recrystallization sections were set to be 270 – 290 °C and 230 – 240 °C, respectively. With this procedure we can obtain platelet-like BP1T-CN crystals with a size of ~100 × 100 $\mu m^2$.

### *Microcavity fabrication*

The bottom DBR was a dielectric thin-film multilayer of $SiO_2$ and $Ta_2O_5$, which was deposited by the rf-magnetron sputtering method on a silica substrate with a thickness of 500 μm. The refractive indices of $SiO_2$ and $Ta_2O_5$ are 1.46 and 2.16, respectively. Deposition of 12 pairs of 80-nm $SiO_2$ and 60-nm $Ta_2O_5$ films results in a high reflectivity larger than 99% at the 440 − 550 nm band. A vapor-grown BP1T-CN crystal was picked up with a thin metallic wire and transferred onto the bottom DBR mirror. The thickness of the crystal was measured to be 365 nm using a contact-type thickness meter (Dektak XT-S, Bruker). The rf-magnetron sputtering method was employed again to deposit a dielectric thin-film multilayer of $SiO_2$ and $HfO_2$ as the top DBR. The refractive index of $HfO_2$ is 2.13. 9.5 pairs of 58-nm $HfO_2$ and 80-nm $SiO_2$ films results in a reflectivity larger than 98 % at the 450 − 575 nm band.

### *Characterizations*

All optical measurements were performed in atmospheric condition (~23 °C and ~40 %RH). We used a cw laser diode (L405P150 & LTC56A/M, Thorlabs) for sample excitation. The excitation wavelength was 405 nm. The emission spectra were measured with a CCD spectrometer (Kymera & Newton, Oxford ANDOR). The emission counts were recorded with time-integration of 1 s and accumulation of 10 times. In ARPL measurements, we used a Fourier space imaging setup using an objective with NA ~ 0.75 (UPLFLN 40X, Olympus). The angle limits of angle resolved measurement is estimated from a relationship of $\theta_{\max} = \sin^{-1}(\mathrm{NA})$ to be 48.6 °. Polarization measurements were performed by using a Glan-Thompson prism and a quarter-wave plate.

*Calculations*

Berreman 4 × 4 matrix method was used for theoretical calculation of $\boldsymbol{k}$-dependent reflectivity spectra from a device that mimics a BP1T-CN microcavity. A thickness of BP1T-CN layer was set to be 365, 730, or 1,095 nm, as explained in the main text. We assumed that the biaxial index ellipsoid of the BP1T-CN crystal consists of (2.15, 1.95, 3.20) as the components in three orthogonal directions. An angle between the optical principal axis and the *xy*-plane was set as 22.5 °, which is almost the same as the experimental fact[24]. The top and bottom DBRs are dielectric multilayers of $SiO_2$ (80 nm)/$HfO_2$ (58 nm) and $SiO_2$ (80 nm)/ $Ta_2O_5$ (60 nm), respectively. The refractive indices of $SiO_2$, $HfO_2$, and $Ta_2O_5$ are 1.46, 2.13, and 2.16, respectively. To adjust to the experimental facts, the numbers of dielectric multilayer layers were reduced to 6.5 and 8 for the top and bottom DBRs, respectively, intentionally degrading the Q factor. To obtain Stokes parameters for reflectivity, a ratio of reflected light power at a certain polarization to the incident power at the same polarization is defined as the reflectance at that polarization.

## Data availability

The datasets generated during and/or analysed during the current study are available from the corresponding author on reasonable request.

## Acknowledgements

This work is supported by Japan Society for the Promotion of Science, JSPS KAKENHI (Nos. 20KK0088, 22K18794, and 22H00215) and from JST CREST (JPMJCR20T4).

## Conflict of interest

The authors declare no conflict of interests.

## Author contributions

Y.Y. and K.Y. conceived and planned the experiments. H.M. grew BP1T-CN crystals using vapor-phase method. M.N. fabricated dielectric thin-film multilayers using magnetron sputtering. R.O. and T.I. performed ARPL measurements. T.I. and K.Y. performed theoretical calculations. R.O., T.I., and K.Y. analyzed experimental and calculated results. S.T. and K.Y. discussed on topological properties. R.O., T.I., and K.Y. drafted the manuscript and complied figures, with discussion of results and feedback from all authors.